\documentclass[b5paper,twoside]{jpconf}  
\usepackage{subfigure}
\usepackage{graphicx}

\begin{document}

\title[Classical and Recurrent Novae]{Classical and Recurrent Nova Outbursts}  

\author[M.F. Bode]{M.F. Bode}

\address{Astrophysics Research Institute, Liverpool John Moores University, Birkenhead, CH41 1LD, UK}

\ead{mfb@astro.livjm.ac.uk} 

\begin{abstract}
Over the last 40 years, multi-frequency observations, coupled with advances in theoretical modelling, have led to a much fuller understanding of the nova phenomenon. Here I give a brief review of the current state of knowledge of Classical and Recurrent Novae including their central systems; the causes and consequences of their outbursts; sub-types, and possible relationships to Type Ia Supernovae. Particular attention is paid to the Recurrent Nova RS Ophiuchi as it shows a wealth of phenomena associated with its 2006 outburst. Finally, some open questions and avenues for future work are summarised. 
\end{abstract}

\section{Introduction}

Nova outbursts have been recorded for over 2000 years (for an historical review, see Duerbeck 2008). It was only in the 1920s during the ``Great Debate''  that it was realised that ``ordinary'' novae such as T Aur (1892 - often seen as the first well-studied nova outburst) were very distinct from ``supernovae'' such as S And (1885 - in M31). Later, Dwarf Novae (DNe) and Classical Novae (CNe) were in turn recognised as rather different beasts and certain of the Classical Novae were also subsequently reclassified as Recurrent Novae (RNe) when a second major outburst was recorded (the earliest example being T Pyx with the 1902 outburst repeating that first noted for this object in 1890. Coincidentally, its latest outburst was 2011 April).

The watershed in our understanding of the outbursts of CNe (and later RNe) came from a combination of the realisation that cataclysmic variables in general, and CNe in particular, had close binary central systems, together with the suggestion that explosive hydrogen burning was the root cause of the outburst (Kraft 1964). The relationship between DNe, CNe and RNe has long been established. The debate over the relationship of RNe in particular to supernovae continues, as discussed below.

\section{Classical Novae}

Historically, the primary defining characteristic of the nova outburst is the optical light curve (see e.g. Strope, Schaefer \& Henden 2010 and Fig.1). Canonically, a rapid rise to maximum is followed by an early decline whose rate defines the ``speed class'' of the nova, with the very fast CNe declining at $> 0.2$ mag d$^{-1}$ (see e.g. Warner 2008). The speed class is in turn related to both the ejection velocity and peak absolute magnitude (in the sense that faster novae have higher ejection velocities and are intrinsically brighter at the peak in their $V$ light output). These properties are of course related to fundamental parameters such as the mass of the accreting white dwarf, and the latter lies behind the Maximum Magnitude-Rate of Decline (MMRD) relations which in the past have been proposed as making CNe potentially important distance indicators. The most widely used MMRD relation is that of Della Valle \& Livio (1995), but this is still subject to significant scatter and uncertainties and is now mainly used as a distance estimator to individual Galactic CNe. The evolution of the CN light curve is also mirrored by a well defined sequence of spectral development comprising in turn the pre-maximum, principal, diffuse-enhanced, Orion, nebular and post-nova spectral stages (see e.g. Duerbeck 2008, Warner 2008).

\begin{figure}
\centering
\mbox{\subfigure{\includegraphics[width=2.5in]{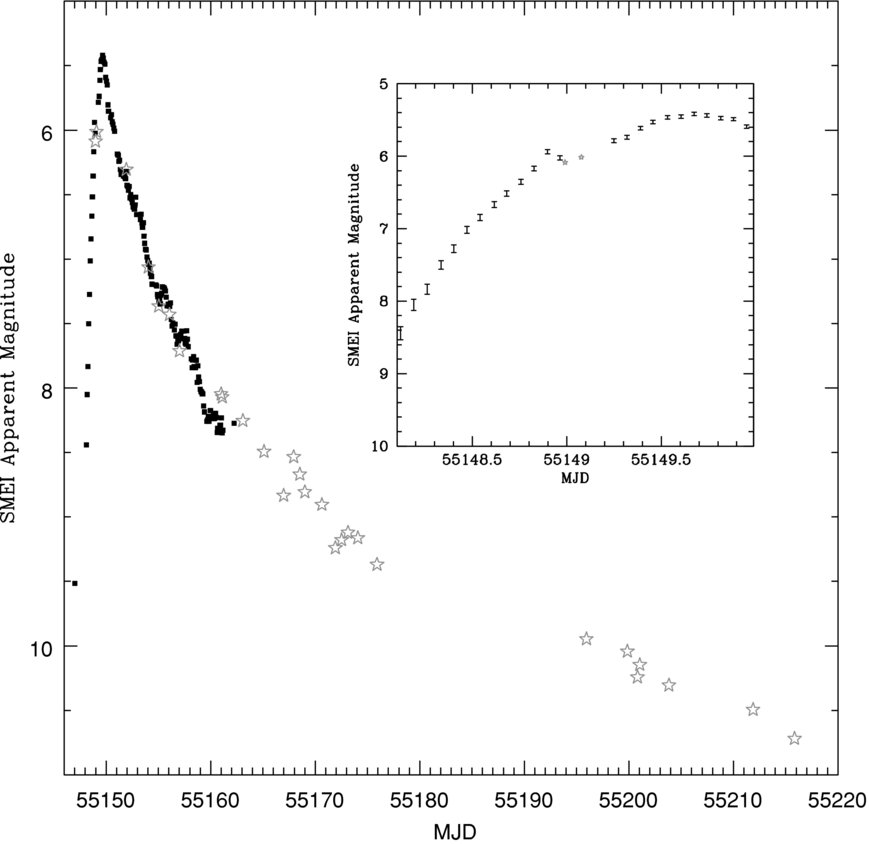}}\quad
\subfigure{\includegraphics[width=2.5in]{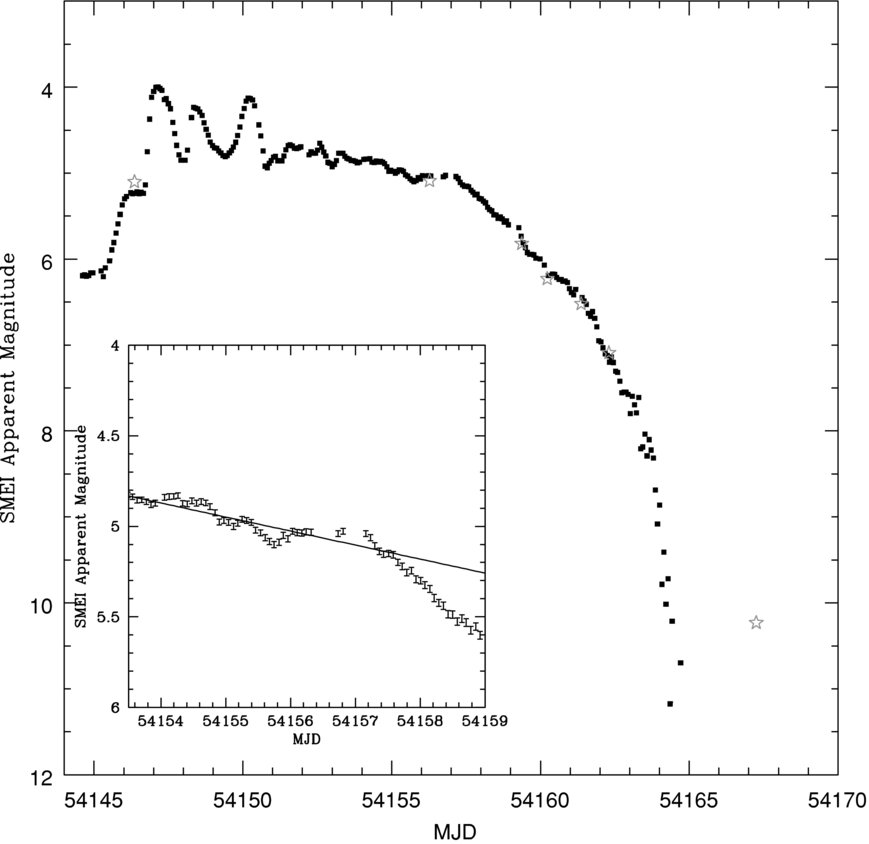} }}
\caption{Unprecedented optical light curves of recent novae with the Solar Mass Ejection Imager (SMEI) aboard the Coriolis satellite. {\em Left:} SMEI (black squares) light curve of the very fast nova KT Eri (2009) with Liverpool Telescope SkyCamT data superimposed (grey stars). The inset shows the rising portion of the light curve with an expanded time scale, clearly demonstrating the existence of a pre-maximum halt. Although the nova reached naked eye visibility at peak, it was not discovered until around 11 days post-maximum. {\em Right:} SMEI light curve of the very slow nova V1280 Sco (2007; black squares), superimposed (grey stars) are data from the ``$\pi$ of the Sky" project. The difference between the late-time SMEI light curve and the $\pi$ of the Sky point may reflect the uncertainties with SMEI data at such faint magnitudes. The inset shows the region around the light curve break which is associated with the onset of dust formation. The solid line shows the fit to the pre-break SMEI light curve and its extrapolation (see Hounsell et al. (2010) for further details).} \label{fig1}
\end{figure}

The central binary comprises a WD primary, of either the CO or ONe type, accreting material from its lower mass companion. The lowest $M_{\rm WD}$ observed in a CN system is $\sim 0.5$M$_\odot$, although selection effects come into play here. The boundary of initial mass between the two WD types is around 1.1 - 1.2 M$_\odot$.  For the observed $P_{\rm {orb}}$ range of  $\sim 1.4 - 8$ hours, the secondary star is a low mass, main sequence object. There are a few longer period systems (e.g. GK Per), but for these, the secondary has to be evolved if it is still to fill its Roche Lobe. The typical accretion rate through the disk is $\dot{M}_{acc}  \sim10^{-9}$ M$_\odot$ yr$^{-1}$ and the luminosity of the quiescent system is $\sim$L$_\odot$ (see e.g. Warner 2008 and references therein).

The CN explosion is due to a thermonuclear runaway (TNR) on the WD surface once sufficient H-rich matter has been accreted at a given WD mass for the critical pressure to be attained at the base of the accreted envelope (there are also effects of composition and the rate of accretion itself to be taken into account, and detailed modelling of the TNR relies on accurate determination of nuclear reaction rates - see e.g. Starrfield, Iliadis \& Hix  2008). It is clear nevertheless that TNR models have been very successful in modelling the general form of the luminosity evolution and observed abundances in novae. However, there are still some unsolved details (see below and e.g Jos\'{e} \& Shore 2008). 

The outburst then results in an increase in luminosity typically to $L \sim$few$\times10^4$L$_\odot$ (i.e. approximately at or above L$_{\rm Edd}$ for a 1 M$_\odot$ WD). As fast novae are inherently more luminous, they are more likely to arise from higher mass WDs. Large amounts of mass are ejected at high velocities typically with $M_{\rm ej}$  $\sim 10^{-5}$ -- few$\times10^{-4}$M$_\odot$ and $v_{\rm ej} \sim$ few$\times10^2$ -- few$\times10^3$ km s$^{-1}$. The heavy element enrichments seen in the ejecta are evidence of mixing between the accreted and the core material. Indeed, CNe are predicted to be the major source of several isotopes in the Galaxy, including $^{15}$N and $^{17}$O (Jos\'{e} \& Hernanz 1998). The inter-outburst period for CN explosions is then thought to be $\sim 10^3 - 10^5$ years, and each system may undergo thousands of outbursts.

As noted below, CNe have been divided into various sub-types, but one often used derives from the observed apparent abundances in the ejecta derived from spectroscopy early in the emission line stage. As noted by Shore (2008) the $CO$ novae show enhanced CNO but generally not heavier elements, and the He/H abundance is near solar. They show a diverse range of light curve behaviour and spectroscopic development and can be prolific dust producers. The $ONe$ or $neon$ novae on the other hand are principally distinguished by the early appearance of the [Ne II] 12.8$\mu$m emission line and often show evidence of depletion of C relative to solar, with associated enhancements of O and Ne. Gehrz (2008) suggests that the most extreme neon novae seem to have much higher ejecta masses than their extreme CO counterparts. As noted by Shore (2008), given the expected rarity of massive progenitors, there has been a surprising number of recent novae of the neon subclass (although we need to be aware of selection effects here of course). Unlike the CO novae, this subclass appears surprisingly uniform in its characteristics.

The optical light curve is however a misleading indicator of the evolution of the bolometric luminosity of a CN outburst. The existence of a constant bolometric phase of post-outburst development was first proposed following multi-frequency observations of FH Ser (1970 - see e.g. Gallagher \& Starrfield 1976). The initial optical decline is then due to a redistribution of flux to shorter wavelengths due to a decreasing mass-loss rate giving rise to a shrinkage of the effective (pseudo)photosphere at constant bolometric luminosity as steady nuclear burning continues on the WD surface. From simple considerations, Bath \& Harkness (1989) find that the effective photospheric temperature is given by $T_{\rm eff} = {\rm T_0} \times 10^{\Delta V / 2.5}$ K, where $\Delta V$ is the decline in $V$ magnitude from peak and T$_{\rm 0}$ = 8000 K (Evans et al. 2005). This relationship does not take into account contributions from emission lines which may be particularly important at later times, and the maximum effective temperature is reached when the pseudo-photosphere has the same radius as that of the WD. Nevertheless, as Bath \& Harkness note, the explicit dependence of the physical conditions on the decline stage is consistent with the observed relationship of the optical decline to the spectral development noted above. A revolution in our understanding of the detailed progress of the outburst is now coming from observations of the Super-Soft Source (SSS) in many novae, largely due to the Swift satellite (see e.g. Schwarz et al. 2011 and below).

In FH Ser, and subsequently several other novae, a visual light curve break and coincident infrared rise are now understood to be due to the formation of an optically thick dust shell, covering the whole of the sky as seen from the nova, and which seems to occur preferentially in moderate speed class novae (see e.g. Evans \& Rawlings 2008, Gehrz 2008, who emphasise the importance of novae as real-time laboratories for the investigation of the formation of astrophysical dusts). Details of the process are still not fully understood however. For example, Shafter et al. (2010) find a strong correlation between speed class and the time of onset of dust condensation that has yet to be explained. Finally, around 20 CNe show relatively slowly evolving radio flux arising from thermal (free-free) emission from the expanding ionised gas shell which is initially optically thick and from which the most reliable estimates of ejected mass are derived (Seaquist \& Bode 2008). Very detailed EVLA monitoring of V1723 Aql by Krauss et al. (2011) has however shown that its radio emission is hard to explain straightforwardly using existing models.

Over 40 CNe now have optically resolved remnants, and around a quarter that number have been resolved in the radio (e.g. O'Brien \& Bode 2008). Optical spectroscopy of  such resolved remnants has allowed their true geometry to be explored. This in turn has led to detailed modelling of the effects on remnant shaping by the early post-outburst common-envelope phase and has also allowed the refinement of expansion parallax distances with consequent potential improvement to the MMRD (see Bode 2002 and references therein). Observations of extended structure in at least two CNe are however most likely related to their pre-nova evolution.

We now know of just under 400 Galactic CNe. Until recent years, the mean detected rate was $\sim$3 per year but now the observed rate is nearer 8 per year as more systematic and deeper sky survey work is undertaken (Pietsch 2010). However, Liller \& Meyer (1987) estimated that the observable number of CNe with $V > 11$ is around 12 per year, and Warner (2008) suggests that a significant number of bright novae are still overlooked (as emphasised by the work of Hounsell et al. 2010; see also Fig. 1). The discovery rate compares to the total estimated Galactic nova rate of $34^{+15}_{-12}$ yr$^{-1}$ deduced by Darnley et al. (2006) from comparison with the rate of $65^{+16}_{-15}$ yr$^{-1}$ they determined for M31 (where around 800 nova candidates have been catalogued to date -  Pietsch 2010).

In terms of the CN population and its relation to the underlying stellar population, Della Valle et al. (1992) found that fast novae are confined to within $z < 100$ pc of the Galactic plane, whereas slow novae range up to $z \sim 1000$ pc (i.e. the Galactic bulge). They suggested that this may be indicative of two distinct classes of nova progenitor. It is indeed natural to imagine higher mass WDs being more common in the disk than in the bulge and therefore giving rise to a higher incidence of fast nova outbursts as a result. Earlier, Arp (1956) had noted a bimodal distribution of speed classes for novae observed in M31 (although Shafter et al. (2009) have cast some doubt on its reality). Williams (1992) has in turn defined two classes of novae based on their outburst spectra: ``FeII novae'' ($\sim60$\% of the observed total) being slower novae having lower ejection velocities, and ``He/N novae'' being faster novae with higher ejection velocities and encompassing the neon novae.

\section{Recurrent Novae}

Unlike CNe, RNe have typical inter-outburst timescales of order decades. 
At present we have secure identifications of a total of only 10 RNe in the Galaxy, although several others are identified in extragalactic systems. Anupama (2008) has attempted to classify the Galactic RNe by sub-type, {\em viz.}: \\ The {\em RS Oph/T
  CrB} group  with red giant secondaries, consequent long orbital periods
($\sim$ several hundred days), rapid declines from outburst ($\sim
0.3$ mag day$^{-1}$), high initial ejection velocities (in excess of 4000 km
s$^{-1}$) and strong evidence for the interaction of the ejecta with
the pre-existing circumstellar wind of the red giant (see next section).\\  The
more heterogeneous {\em U Sco} group with members' central systems
containing an evolved main sequence or sub-giant secondary with an
orbital period much more similar to that in CNe (of order hours to a
day), rapid optical declines (U Sco itself being one of the fastest
declining novae of any type), extremely high ejection velocities
($v_{ej} \sim 10,000$ km s$^{-1}$, from FWZI of emission lines for U
Sco) but no evidence of the extent of shock interactions seen in RS Oph
following outburst (their post-outburst optical spectra resemble the ``He/N''
class of CNe).\\ The {\em T Pyx} group
again comprises short orbital period systems and although their optical spectral
evolution following outburst is similar, with their early-time spectra
resembling the ``Fe II'' CNe, they show a very heterogeneous set of
moderately fast to slow optical light curve declines. This group of systems also seems to show ejected masses similar to
those at the lower end of the ejected mass range for CNe with $M_{\rm
  ej} \sim 10^{-5}$ M$_{\odot}$ (i.e. one to two orders of magnitude
greater than $M_{\rm ej} $ in the other two sub-types of RNe).

It is interesting to note that we have witnessed the outbursts of all three of the prototype systems for each sub-type in the last 5 years ({\em viz.} RS Oph, 2006; U Sco, 2010, and T Pyx, 2011) and these outbursts are again due to TNR on a WD dwarf surface. The short recurrence periods of RNe require high mass WD accretors and
relatively high accretion rates (e.g. Starrfield et al. 1988, Yaron et al. 2005). Indeed, both
RS Oph and U Sco appear to have WDs near to the Chandrasekhar mass
limit. Unlike CNe where the WD mass is generally thought  to be decreasing (although this is still an interesting open question in fact), the WD mass in both these objects has been proposed as growing
such that they are potential SN Ia progenitors (see e.g. Osborne et al. 2011 and Kahabka et al. 1999 respectively). The study of RNe is thus important for several broader fields of
investigation including mass loss from red giants, the evolution of
SNR and the determination of the progenitors of Type Ia SNe.

\section{The Recurrent Nova RS Ophiuchi}

We pay particular attention to this RN as its 2006 outburst has been extremely thoroughly observed and it shows a wealth of phenomena. RS Oph has had recorded outbursts  in 1898, 1933, 1958,1967, 1985 and 2006, plus probable eruptions in 1907 and 1945. The optical behaviour from one outburst to the next is very similar. The central system comprises a high mass WD in a 455 day orbit with a red giant (M2III) and $d$ (= 1.6$\pm 0.3$ kpc) and $N_{\rm H}$ ($= 2.4\pm0.6 \times 10^{21}$ cm$^{-2}$) are well defined (see Evans et al. 2008, and papers therein).

Optical spectra from outbursts prior to 1985 showed evidence for the existence of a low velocity red giant wind ($u \simeq 20$ km s$^{-1}$) into which the high velocity ejecta ($v_{\rm 0}  \simeq 4000$ km s$^{-1}$) were running. The 1985 outburst was the first to be observed beyond the visual but it was only with the latest eruption on 2006 February 12 that very detailed radio imagery and X-ray observations in particular could be performed. The X-ray evolution was observed from around 3 days after outburst by both Swift (Bode et al. 2006) and RXTE (Sokoloski et al. 2006). These observations revealed a bright, rapidly evolving thermal source consistent with emission from the shocked wind ahead of the ejecta. For Swift, the source was clearly detected in the lowest energy channel of the BAT (14-25 keV) at the time of the outburst itself and source evolution was followed for around 200 days with the XRT (e.g. Page et al. 2008, Bode et al. 2008).

The XRT spectra obtained over approximately the first three weeks post-outburst could be well fitted with a single temperature {\em mekal} model with the interstellar $N_{\rm H}$ fixed and that from the overlying wind allowed to be a free parameter. The shock velocity was derived from the temperature resulting from the fit. The results were consistent with a rapid change at around 6 days after outburst from ``Phase I'' to ``Phase III'' behaviour (Bode et al. 2006) as defined for supernova remnants where the forward shock moves into a $1/r^2$ density distribution (Bode \& Kahn 1985). Subsequently, Nelson et al. (2008), Drake et al. (2009) and Ness et al. (2009) have presented the results of their analyses of grating observations with Chandra and XMM-Newton, which commenced 13.8 days after outburst. These have allowed them to explore in more detail such things as shock evolution and elemental abundances in the emitting gas. Vaytet et al. (2011) also present detailed hydrodynamical models of the evolution of the X-ray emission observed in the Swift XRT.

At around 26 days a new, low energy, spectral component rapidly emerged which was initially highly variable and dominated the X-ray emission for approximately the next 70 days (Osborne et al. 2011). Indeed, most of the additional count rate resulted from a huge increase in counts below 0.7keV. Here again, Ness et al. (2007) and Nelson et al. (2008) have presented analyses of Chandra and XMM-Newton spectroscopy at several epochs during this SSS phase. The duration of the SSS phase was relatively short, implying that the mass of the WD is high (consistent with the short inter-outburst period - see also Hachisu et al. 2007) and a 35s quasi-periodic modulation was present up to day 65 (Beardmore et al. 2008). The origin of this very short period modulation is still enigmatic, but remarkably, this has subsequently been seen in nova KT Eri (2009; Beardmore et al. 2010). A much longer period oscillation has been observed in V1494 Aql (Drake et al. 2003) and proposed as due to non-radial g$^+$ modes from  the WD. The 35s pulsation seen in RS Oph may however be due to a nuclear burning instability similar to that suggested for (albeit longer period) oscillations in a PN nucleus (Kawaler 1988).

\begin{figure}
\begin{center}
\includegraphics[width=100mm,height=69.4mm]{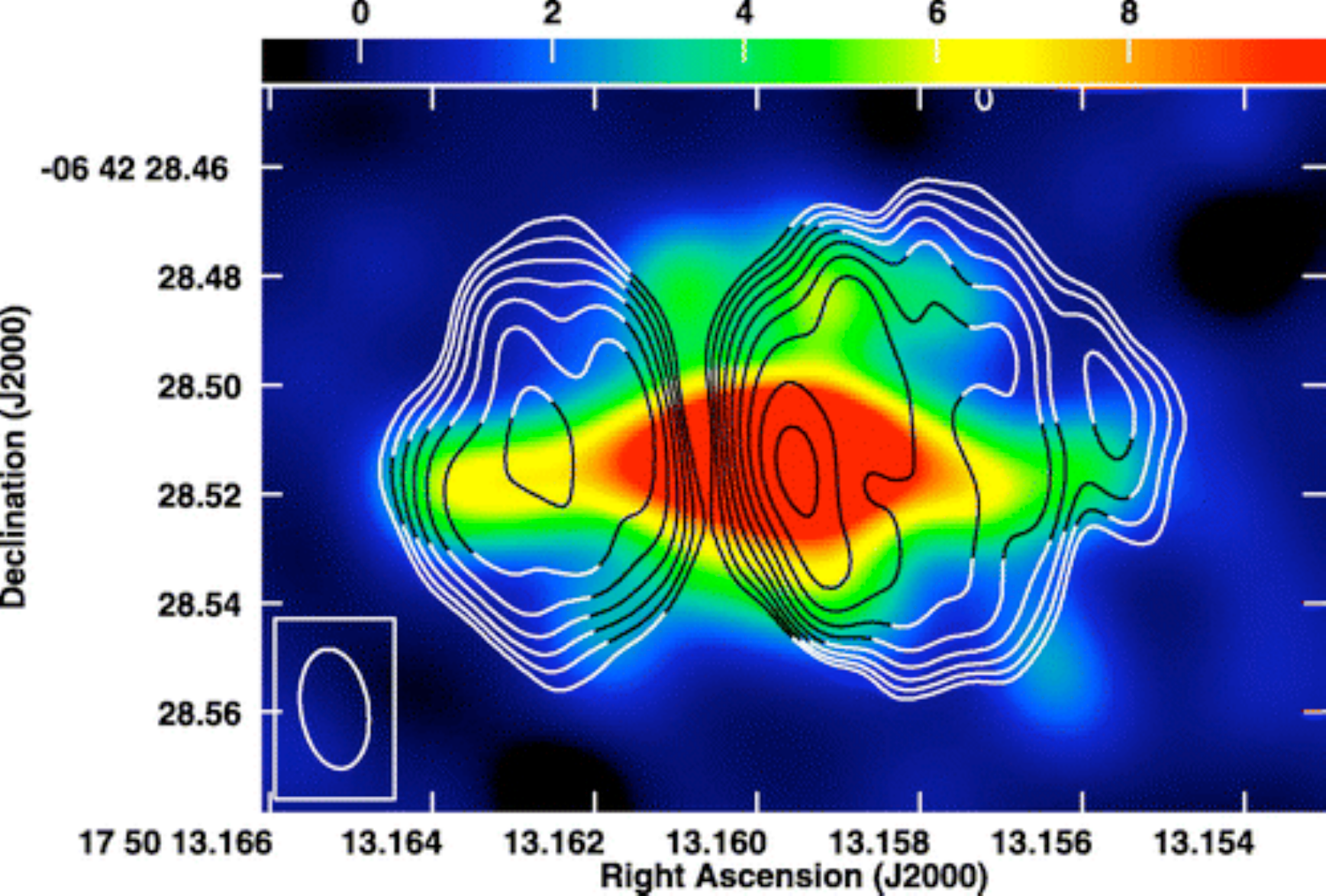}
\caption{Highly collimated outflows and lobes in RS Oph around 50 days from outburst. The contour lines show 1.7 GHz emission (as imaged by the VLBA), which preferentially traces synchrotron-emitting gas. The grey-scale image shows 43 GHz emission (VLA), which preferentially traces thermal plasma. North is up and east is to the left (from Sokoloski et al. 2008).}
\label{label4}
\end{center}
\end{figure}

Turning now to the radio, VLA and MERLIN observations began within a few days of the 2006 outburst (Eyres et al. 2009). VLBI imaging started on day 13 with the VLBA resolving a partial ring of non-thermally dominated emission of radius around 9 mas, consistent with the dimensions expected for the forward shock at this time as derived from modelling of the X-ray emission (O'Brien et al. 2006, see also Rupen et al. 2008). More extended emission in the form of two lobes gradually appeared to the E then the W and the whole structure eventually bore a close resemblance to that derived from comparatively rudimentary VLBI observations secured by Taylor et al. (1989) on day 77 of the 1985 outburst (Sokoloski et al. 2008; it should also be noted that coincidentally the binary phase was approximately the same at both epochs). Indeed, the suggestion from the 1985 observations that there was a thermal core of emission centred between two non-thermal lobes was borne out by observations of the 2006 outburst (see Fig. 2). Sokoloski et al. (2008) went on to suggest that there were two jets of emission emanating from the central system, directed E-W, and giving rise to non-thermal lobes at the working surfaces of the jets (again, see Fig. 2).

Hubble Space Telescope observations of the expanding nebular remnant were conducted on days 155 and 449 through narrow band filters. Extended structure was particularly evident in [OIII]$\lambda5007$ where two lobes of emission were clearly present (Bode et al. 2007). The spatial extent of the nebula was consistent with that in the radio with expansion velocity $\simeq 3200$ km s$^{-1}$ in the plane of the sky, and no deceleration of the outermost parts apparent between the two epochs. The combination of contemporaneous ground-based optical spectroscopy and modelling of the day 155 image led Ribeiro et al. (2009) to conclude that the lobes arise from a dumbbell-shaped emission region and that there is a region of much higher density material in the centre which in contrast may show some evidence for deceleration. Furthermore, they were able to determine the inclination of the nebula to the plane of the sky as $39^{+1}_{-10}$ degrees and that the West lobe is approaching the observer (in line with conclusions from early-time VLTI infrared interferometry by Chesneau et al. 2007). The true maximum expansion velocities then exceed 5000 km s$^{-1}$.

Bode et al. (2007) and Ribeiro et al. (2009) suggest that the axis of the lobes lies along that of the binary orbit and that there is an enhancement in the density of the pre-existing red giant wind in the equatorial plane of the binary (as has been proposed for such systems by e.g. Mastrodemos \& Morris 1999). Shock propagation is therefore much more rapid in the direction of the orbital pole than in the plane (see also Walder et al. 2008; Drake et al. 2009; Orlando et al. 2009). Whether the collimation of any jet also occurs because of this anisotropic distribution of circumbinary material, or is due to interaction with an accretion disk around the WD, is an open question. The existence of a dense region near the centre of the system which shocks may not entirely traverse would also explain the unexpected survival of significant amounts of dust following the outburst, as revealed by Spitzer (Evans et al. 2007).

\section{Concluding remarks and  {\em some} open questions}

\begin{itemize}

\item{Is there a continuum of inter-outburst timescales (and other fundamental properties) from CNe through RNe (at least for the short period sub-types of RNe)? Studies of large samples of novae in extragalactic systems such as M31 are likely to be most help here.}

\item{What is the cause of the remarkable variability during the emergence of the SSS in RS Oph and other novae, and also, is a nuclear burning instability really the origin of the 35s modulation seen in the XRT data for both RS Oph and KT Eri while the SSS was bright?}

\item{There is still a significant (order of magnitude) discrepancy between the ejected mass found from models of the CN outburst compared to that derived from observations (although this seems less of a problem with the U Sco and RS Oph sub-classes of RNe). Consideration of clumpiness in the ejecta helps to a degree, but does not appear sufficient and it may mean we need to re-examine our models.}

\item{Some details of the grain formation process remain problematic; we don't understand for example the correlation between nova speed class and grain formation timescale.}

\item{Can studies of objects such as RS Oph help to determine more precisely the mechanism of the formation of jets in astrophysical sources? Although examples among the nova population may be rare, and perhaps confined only to the RNe, there is no doubting the wider importance of such work.}

\item{Is there indeed a link between RNe and Type Ia SNe? Although it seems that for systems like RS Oph and U Sco, the mass of the WD is near the Chandrasekhar limit and increasing, there remain some fundamental questions. For example, what is the type of the WD (if ONe, then no SN explosion will occur - this has been suggested in the case of U Sco by Mason, 2011); can the H in systems with red giant secondaries really be ``hidden'' at the time of any SN outburst (we note also that Li et al. (2011) rule out an RS Oph-like system as the progenitor of SN 2011fe in M101); is the population of (appropriate sub-type) RNe sufficient to explain the observed SN Ia rate? A combination of detailed observations of individual Galactic RNe,  coupled with surveys of extragalactic novae, will help to answer these very important points.}

\item{How many bright Galactic novae are we still missing? The SMEI results of Hounsell et al. (2010) are instructive in this regard and the exploration of the full SMEI dataset should help resolve this important issue.}

\end{itemize}

\section*{References}
\begin{thereferences} 
\item Anupama, G.C.,  2008, in RS Ophiuchi (2006) and the Recurrent Nova Phenomenon, edited by A. Evans, M.F. Bode, T.J. O'Brien and M.J. Darnley, ASP Conf Series, 401, 31
\item Arp, H.C.,  1956, AJ 61, 15
\item Bath, G.T., \& Harkness, R.P.,  1989, in Classical Novae, 1st Ed., edited by M.F. Bode and A. Evans, Wiley, p61
\item Beardmore, A.P.,  Osborne, J.P., Page, K.L., Goad, M.R., Bode, M.F., \& Starrfield, S.,  2008, in RS Ophiuchi (2006) and the Recurrent Nova Phenomenon, edited by A. Evans, M.F. Bode, T.J. O'Brien and M.J. Darnley, ASP Conf Series, 401, 296
\item Beardmore, A.P., et al.,  2010, ATel 2423
\item Bode, M.F., \& Kahn, F.D.,  1985, MNRAS 217, 205
\item Bode, M.F.,  2002, in Classical Nova Explosions, edited by M. Hernanz and J. Jos$\acute{e}$, AIP Conf Proceedings, p497
\item Bode, M.F., et al.,  2006, ApJ 652, 629
\item Bode, M.F., et al., 2007, ApJ 665, L63
\item Bode, M.F., et al., 2008, in RS Ophiuchi (2006) and the Recurrent Nova Phenomenon, edited by A. Evans, M.F. Bode, T.J. O'Brien and M.J. Darnley, ASP Conf Series, 401, 269
\item Chesneau, O., et al.,  2007, A\&A 464, 119
\item Darnley, M.J., et al.,  2006, MNRAS 369,257
\item Della Valle, M., Bianchini, A., Livio, M., \& Orio, M.,  1992, A\&A 266, 232
\item Della Valle, M., \& Livio, M.,  1995, ApJ 452, 704
\item Drake, J.J., et al: 2003, ApJ 584, 448
\item Drake, J.J., et al: 2009, ApJ 691, 418 
\item Duerbeck, H.R.,  2008, in Classical Novae, 2nd Ed., edited by M.F. Bode and A. Evans, CUP, p1
\item Eyres, S.P.S., et al.,  2009, MNRAS, 395, 1533
\item Evans, A., Tyne, V.H., Smith, O., Geballe, T.R., Rawlings, J,M.C., \& Eyres, S.P.S.,  2005, MNRAS 360, 1483
\item Evans, A., et al.,  2007, ApJ 671, L157
\item Evans, A., \& Rawlings, J.M.C.,  2008, in Classical Novae, 2nd Ed., edited by M.F. Bode and A. Evans, CUP, p308
\item Evans, A., Bode, M.F., O'Brien T.J., \& Darnley, M.J. (editors), 2008, RS Ophiuchi (2006) and the Recurrent Nova Phenomenon,  ASP Conf Series
\item Gallagher, J.S, \& Starrfield, S.,  1976, MNRAS 176, 53
\item Gehrz, R.D.,  2008, in Classical Novae, 2nd Ed., edited by M.F. Bode and A. Evans, CUP, p167
\item Hachisu, I., Kato, M., \& Luna, G.J.M.,  2007, ApJ 659, L153
\item Hounsell, R., et al., 2010, ApJ, 724, 480
\item Jos\'{e}, J.,  \& Hernanz, M.,  1998, ApJ, 494, 680
\item Jos\'{e}, J.,  \& Shore, S.N,. 2008, in Classical Novae, 2nd Ed., edited by M.F. Bode and A. Evans, CUP, p121
\item Kahabka, P., Hartmann, H.W., Parmar, A.N., \& Negueruela, I.,  1999, A\&A 347, L43
\item Kawaler, S.D.,  1988, ApJ 334, 220
\item Kraft, R.P.,  1964, ApJ 139, 457
\item Krauss, M.I. et al.,  2011, ApJ, 739, L6
\item Li, W., et al.,  2011, Nature, in press; arXiv:1109.1593
\item Liller, W., \& Mayer, B.,  1987, PASP 99, 606
\item Mason, E.,  2011, A\&A, 532, L11
\item Mastrodemos, N., \& Morris, M.,  1999, ApJ 523, 357
\item Nelson, T., Orio, M., Cassinelli, J.P., Still, M., Leibowitz, E., \& Mucciarelli, P.,  2008, ApJ 673, 1067
\item Ness, J-U, et al.,  2007, ApJ 665, 1334
\item Ness, J-U, et al.,  2009, AJ 137, 3414
\item O'Brien, T.J., Lloyd, H.M., \& Bode, M.F.,  1994, MNRAS 271, 155
\item O'Brien, T.J., \& Bode, M.F.,  2008, in Classical Novae, 2nd Ed., edited by M.F. Bode and A. Evans, CUP, p285
\item O'Brien, T.J., et al.,  2006, Nature 442,279
\item Orlando, S., Drake, J.J., \& Laming, J.M.,  2009, A\&A 493, 1049
\item Osborne, J.P. et al.,  2011, ApJ, 727, 124
\item Page, K.L., 
et al.,  2008, in RS Ophiuchi (2006) and the Recurrent Nova Phenomenon, edited by A. Evans, M.F. Bode, T.J. O'Brien and M.J. Darnley, ASP Conf Series, 401, 283
\item Pietsch, W., 2010, AN, 331, 187
\item Ribeiro, V.A.R.M., et al., 2009, ApJ 703, 1955
\item Rupen, M.P., Mioduszewski, A.J., \& Sokoloski, J.L.,  2008, ApJ 688, 559
\item Schwarz, G.J., et al.,  2011, ApJS, in press; arXiv:1110.6224
\item Seaquist, E.R., \&. Bode, M.F.,  2008, in Classical Novae, 2nd Ed., edited by M.F. Bode and A. Evans, CUP, p141
\item Shafter, A.W., Rau, A., Quimby, R.M., Kasliwal, M.M., Bode, M.F., Darnley, M.J., \& Misselt, K.A.,  2009, ApJ 690, 1148
\item Shafter, A. W., Bode, M. F., Darnley, M. J., Misselt, K. A., Rubin, M., \& Hornoch, K.,  2010, ApJ, 734, 12 
\item Shore, S.N.,  2008, in, Classical Novae, 2nd Ed., edited by M.F. Bode and A. Evans, CUP, 194
\item Shore, S.N., Starrfield, S., Ake, T.B., \& Hauschildt, P.H.,  1997, ApJ 490, 393
\item Sokoloski, J.L., Luna, G.J M., Mukai, K., \& Kenyon, S.J.,  2006, Nature 442, 276
\item Sokoloski, J.L., Rupen, M.P.,  \& Mioduszewski, A.J.,  2008, ApJ 685, L137
\item Starrfield, S., Sparks, W.M., \& Shaviv, G., 1988, ApJ 325, L35 
\item Starrfield, S., Truran, J.W., Wiescher, M.C., \& Sparks, W.M.,  1998, MNRAS 296, 502
\item Starrfield, S., Iliadis, C., \& Hix, W.R.,  2008, in, Classical Novae, 2nd Ed., edited by M.F. Bode and A. Evans, CUP, p77
\item Strope, R.J., Schaefer, B.E.,  \& Henden, A.A., 2010, AJ 140, 34
\item Taylor, A.R., Davis, R.J., Porcas, R.W.,  Bode, M.F.,  1989, MNRAS 237, 81
\item Vaytet, N.M.H., O'Brien, T.J., Page, K.L., Bode, M.F., Lloyd, M., \& Beardmore, A.P. 2011, ApJ, 740, 5
\item Walder, R., Folini, D., \& Shore, S.N.,  2008, A\&A 484, L9
\item Warner, B.,  2008, in Classical Novae, 2nd Ed., edited by M.F. Bode and A. Evans, CUP, p16
\item Williams, R.E.,  1992, AJ 104, 725
\item Yaron, O., Prialnik, D., Shara, M. M., \& Kovetz, A.,  2005, ApJ 623, 398

\end{thereferences}

\end{document}